\documentclass[reprint,secnumarabic,amssymb, nobibnotes, aps, prd]{revtex4-1}
\usepackage{hyperref}
\usepackage{graphicx}

\newcommand{\Ho}{$^{163}$Ho}
\newcommand{\Dy}{$^{163}$Dy}
\newcommand{\QEC}{$Q_{\mathrm{EC}}$}

\begin{document}

\title{First Calorimetric Measurement of OI-line in the Electron Capture Spectrum of \Ho}%

\author{P. C.-O. Ranitzsch}
\email[P. C.-O. Ranitzsch: ]{philipp.ranitzsch@kip.uni-heidelberg.de}
\author{C. Hassel}
\author{M. Wegner}
\author{S. Kempf}
\author{A. Fleischmann}
\author{C. Enss}
\author{L. Gastaldo}
\affiliation{Kirchhoff-Institute for Physics, Im Neuenheimer Feld 227, 69120 Heidelberg, Germany}
\author{A. Herlert} \altaffiliation{Present address: FAIR GmbH, Planckstr. 1, D-64291 Darmstadt, Germany}
\author{K. Johnston} \altaffiliation{Also: Technische Physik, Universit\"at des Saarlandes, 66041 Saarbr\"ucken, Germany}
\affiliation{CERN, Physics Department, 1211 Geneva 23, Switzerland}
\date{August 26, 2014}%

\begin{abstract}

The isotope \Ho\ undergoes an electron capture process with a recommended value for the energy available to the decay, \QEC, of about $2.5\ \mathrm{keV}$. According to the present knowledge, this is the lowest \QEC\ value for electron capture processes. Because of that, \Ho\ is the best candidate to perform experiments to investigate the value of the electron neutrino mass based on the analysis of the calorimetrically measured spectrum. 

We present for the first time the calorimetric measurement of the atomic de-excitation of the \Dy\ daughter atom upon the capture of an electron from the $5s$ shell in \Ho, OI-line. The measured peak energy is $48\ \mathrm{eV}$. This measurement was performed using low temperature metallic magnetic calorimeters with the \Ho\ ion implanted in the absorber.

We demonstrate that the calorimetric spectrum of \Ho\ can be measured with high precision and that the parameters describing the spectrum can be learned from the analysis of the data.

Finally, we discuss the implications of this result for the Electron Capture \Ho\ experiment, ECHo, aiming to reach sub-eV sensitivity on the electron neutrino mass by a high precision and high statistics calorimetric measurement of the \Ho\ spectrum. 

\end{abstract}

\maketitle
\tableofcontents

\section{Introduction}

The isotope \Ho\ is among the known nuclides, undergoing electron capture (EC) processes, the one with the lowest energy available to the decay. The recommended value is $Q_{\mathrm{EC}}=(2.555\pm0.016)\ \mathrm{keV}$ \cite{Wang2012}. This property makes the \Ho\ the best candidate for the investigation of the electron neutrino mass in the sub-eV energy region by the kinematical analysis of EC spectrum.

De Rujula and Lusignoli proposed for the first time in 1982 that, in the case of \Ho, it is possible to reach high sensitivity to a non-zero electron neutrino mass if all the energy released in the decay, minus the energy taken away by the electron neutrino, is measured~\cite{DeRujula1982}. In the years following the publication, several experiment showed the possibility to perform the calorimetric measurement of the EC spectrum of \Ho~\cite{Hartmann1985, Andersen1982, Gatti1997}. Nevertheless the achieved performance did not reach the level to make a \Ho-based experiment competitive with the experiment investigating the mass of the electron antineutrino by the kinematical analysis of the beta spectrum of $^{3}$H. Recent measurements performed by our group using low temperature metallic magnetic calorimeters (MMCs) \cite{Fleischmann2009}, showed that it is possible to measure with high precision the EC spectrum of \Ho~\cite{Ranitzsch2012, Gastaldo2013}. These results motivated the formation of the international collaboration ECHo (Electron Capture \Ho) with the aim to investigate the electron neutrino mass in the sub-eV range by means of the high precision and high statistics calorimetric measurement of the EC spectrum of \Ho\ \cite{Gastaldo2014}. The measurements discussed in this paper are performed within the ECHo experiment. Meanwhile also a second similar experiment named HOLMES was funded \cite{HOLMES2013} and several groups are performing R\&D for source production and detector development \cite{Engle2013, Croce2014, Fowler2013}.

 We present the first calorimetric measurement of de-excitation of the \Dy\ daughter atom after the capture of an electron from the $5s$ shell in \Ho.  With this measurement we demonstrate that the calorimetric spectrum of \Ho\ can be measured with high precision and that the parameters describing the spectrum can be learned from the analysis of the data.

In the following we will give a brief description of the detection principle of MMCs and how they can be fabricated for the calorimetric measurement of the EC spectrum of \Ho. We will then discuss the measurement that has led to the detection of the OI-line and the data reduction methods. Finally we will discuss the implication of this measurement for future experiment for the investigation of the electron neutrino mass.

\section{Metallic magnetic calorimeters}

Metallic magnetic micro-calorimeters are energy dispersive detectors operated at temperatures below $100\ \mathrm{mK}$. These detectors can be described as a particle absorber tightly connected to a sensor which is then weakly connected to a thermal bath. The temperature of the detector increases upon the absorption of energy. The temperature sensor is a paramagnetic alloy, usually Au:Er with an erbium concentration of about $100\dots1000\ \mathrm{ppm}$ sitting in an external magnetic field. Any change of temperature leads to a change of magnetization of the sensor which is read out as a change of magnetic flux in a low noise high band-width dc-SQUID \cite{Fleischmann2009}. In order to reach high sensitivity in the soft X-ray region the volume of these detectors is well below $1\ \mathrm{mm^{3}}$. The resolving power $E/\Delta E$ approaching 5000, the intrinsic response time well below $1\ \mathrm{\mu s}$ and the excellent linearity make MMCs very attractive for numerous experiments \cite{Pies2012}.

With these detectors we have performed the first high resolution calorimetric measurements of the \Ho\ electron capture spectrum \cite{Ranitzsch2012, Gastaldo2013}. The chip prototype we have developed for these measurements consists of four single pixel detectors. To perform the calorimetric measurement of the EC spectrum of \Ho, the source had to be embedded in the absorber. This was obtained by implanting the \Ho\ ions, at a depth of about $5\ \mathrm{nm}$, on a $160\times160\ \mathrm{\mu m^{2}}$ surface of a $190\times190\times5\ \mathrm{\mu m^{3}}$ gold layer which represents the first part of the absorber. The implantation was simultaneously performed on the absorber of each of the four pixels. This process was performed at ISOLDE-CERN \cite{ISOLDE2000}. After the implantation, a second gold layer with the same dimensions as the first one was deposited on top of the first one. This sandwich design of the absorber ensured a quantum efficiency close to 100\% for the energy emitted in the EC decay of \Ho.

\section{Experiment}

The data discussed in the following has been acquired by measuring simultaneously two of the four single pixels which are present on the chip prototype. An integrated two-stage SQUID of the type C6X114W produced by the PTB in Berlin \cite{PTB} was used to read-out each of the two pixels. The signal of each detector, after the SQUID electronics XXF-1 FLL \cite{MagniconXXF1} was split into two channels one of them, for the trigger, going through a strong band pass filter and the other being only low-pass filtered to prevent aliasing. Each of the four channels, one trigger and one detectors signal for each pixel were digitalized by a 4-channel 14-bit digitizer \cite{GaGeCSE1442}. Whenever one of the two trigger signals exceeded the trigger level, both detector signals were acquired on the same time window. The synchronization of the data acquisition for the two pixels allowed for the thermal cross-talk analysis for the reduction of background.

\section{Data analysis}

The two detectors used in the measurement are fabricated on the same Si chip. A detailed description is given in~\cite{Gastaldo2013}. Due to the non-gradiometric setup of the temperature sensors, the detectors are also susceptible to temperature changes of the Si substrate. This leads to two classes of events besides the release of energy in the absorber or sensor of the MMC. Firstly, events due the particles releasing their energy in the substrate. This kind of events will lead to thermal signals of about the same amplitude in the two detectors. Secondly, events due to the so called thermal crosstalk which consist in the fact that the energy released in one of the pixels, flows through the substrate to the thermal bath. The slightly increased temperature of the substrate is measured by neighboring pixels. For this detector design,  thermal cross-talk events have an amplitude corresponding to about 1 \% of the  signal in the primary detector. For events in the primary detectors due to the EC in \Ho, the maximum cross-talk energy will be smaller than 30 eV.

\begin{figure}[b]
	\begin{center}	
	\includegraphics[%
	  width=0.75\linewidth,
	  ]{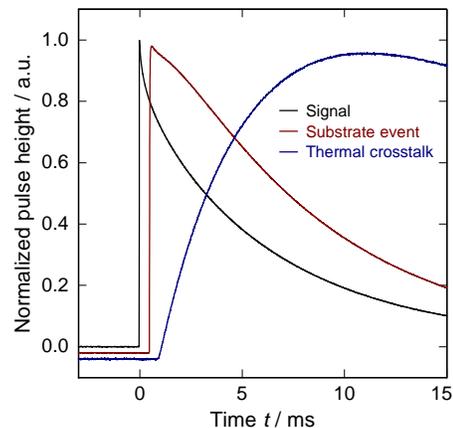}
	\end{center}
	\caption{Exemplary events for the events in the detector and the two different types of background events. 
	\label{fig_Compare_Pulses}
	}
	\end{figure}

These two classes of events are, in respect to \Ho\ spectrum, a source of background. Two characteristic aspects of those events can be used to identify them and eliminate from the spectrum: they lead to simultaneous signals in the two detectors and have in general a different shape compared to direct events.  Figure~\ref{fig_Compare_Pulses} shows the typical shape for direct events, substrate events and crosstalk events. 

\begin{figure}[b]
	\begin{center}	
	\includegraphics[%
	  width=\linewidth,
	  keepaspectratio]{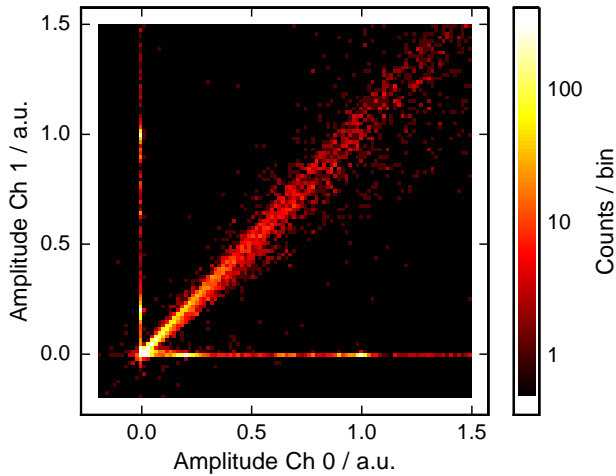}
	\end{center}
	\caption{The amplitudes of the two recorded detector channels plotted against each other, while the event abundance is color coded. The horizontal and vertical lines show the events directly absorbed in the corresponding detector while the diagonal represents the background events.
	\label{fig_Amp_vs_Amp}
	}
	\end{figure}

Using only pulse shape discrimination it is possible to identify  background pulses down to an energy of about 500 eV. For smaller energy the signal to noise ration is not enough to safely distinguish the classes. The synchronization of the data acquisition of the two pixels adds the condition of simultaneity which lowers the energy threshold for separation of the three classes down to a few tens of eV. Figure~\ref{fig_Amp_vs_Amp} shows a scatter plot where the amplitude of channel 1 is plotted versus the ampltude of channel 0. The events showing amplitude = 0 in channel 1 or channel 0 correspond to direct events in the opposing channel, while the events scattering around the diagonal belong to the two background classes just discussed.

\begin{figure*}[t]
	\begin{center}	
	\includegraphics[%
	  width=\linewidth,
	  keepaspectratio]{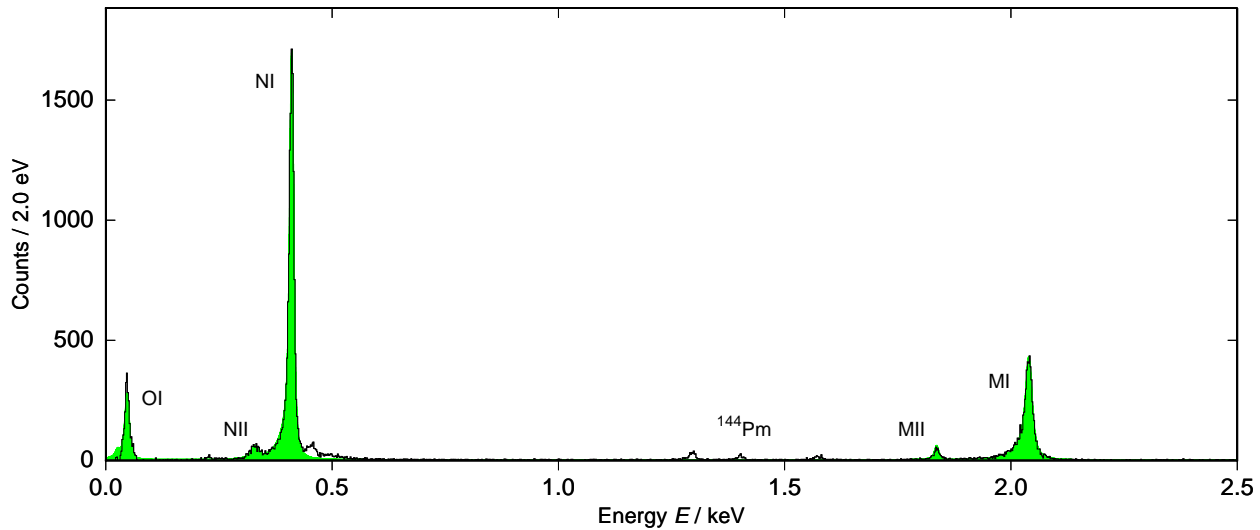}
	\end{center}
	\caption{The measured electron capture spectrum of $^{163}$Ho (black histogram) and the theoretical description (green filled area) with an energy resolution of $\Delta E_{\mathrm{FWHM}}=8.3\,\mathrm{eV}$. Details are in the text and in table~\ref{tab_energies}. }
	\label{fig_Spec_full}
	\end{figure*}

The background pulses can be eliminated down to about $40\ \mathrm{eV}$. By selecting the direct events in channel 0 and channel 1 and summing several of such measurements, the spectrum showed in figure~\ref{fig_Spec_full} is obtained. For the first time it was possible to measure calorimetrically the OI-line corresponding to the capture of a $5s$ electron at $48\ \mathrm{eV}$.

\begin{table}[b]
	\begin{center}
	\begin{tabular}{l||rrrrr}
	
	H & 	MI		& MII		& NI		& NII		& OI	\\ \hline \hline
	$B_{\mathrm{H}}\frac{\varphi_{\mathrm{H}}^2(0)}{\varphi_{\mathrm{MI}}^2(0)}$ \cite{Faessler2014}&%
	1 & 0.051 & 0.244 & 0.012 & 0.032 \\ \hline
	$E_{\mathrm{H}}^{\mathrm{lit}}\ [\mathrm{eV}]$ \cite{Deslattes2003}%
	& 2046.9 & 1844.6 & 420.3 & 340.6 & 49.9~\cite{XRay2009} \\
	$E_{\mathrm{H}}^{\mathrm{exp}}\ [\mathrm{eV}]$%
	& 2040 & 1836 & 411.1 & 330.3 & 48 \\
	$\Gamma_{\mathrm{H}}^{\mathrm{lit}}$ $[\mathrm{eV}]$ \cite{Campbell2001}%
	& 13.2 & 6.0 & 5.4 & 5.3 & 3.7~\cite{Cohen1972} \\
	$\Gamma_{\mathrm{H}}^{\mathrm{exp}}$ $[\mathrm{eV}]$%
	& 13.4 & 4.8 & 4.7 & 13 & 5.6 \\ \hline \hline
	\end{tabular}
	\caption{The parameters going into ($\varphi_{\mathrm{H}}^2(0)$, $B_{\mathrm{H}}$) or coming out of the fit ($E_{\mathrm{H}}^{\mathrm{exp}}$, $\Gamma_{\mathrm{H}}^{\mathrm{exp}}$) to the measured spectrum. The experimental results are compared to their literature values. The literature reference is given in the left column, exceptions are given directly next to the data value.
	\label{tab_energies}
	}
	\end{center}
	\end{table}

This anti-coincidence pulse discrimination leads to an improvement of the low energy threshold by a factor of 5 from 200 eV in previous measurements~\cite{Ranitzsch2012, Gastaldo2014} to 40 eV in the presented spectrum shown in figure~\ref{fig_Spec_full}. This improvement allowed for the first observation of the de-excitation energy of an electron capture from the \Ho\ $5s$ shell, the OI-line, at an energy of $E=48\ \mathrm{eV}$. 
Figure~\ref{fig_Spec_full} shows the spectrum, where the MI-, MII-, NI-, NII- and OI-lines are visible, while details are given in table~\ref{tab_energies}. Besides the lines from the \Ho\ electron capture, also three de-excitation lines of the $^{144}$Pm electron capture can be seen around $1.5\ \mathrm{keV}$. This impurity was ion-implanted as $^{144}$PmF$^+$ alongside the desired \Ho.

Following~\cite{DeRujula1982} the differential spectrum can be described by:

\begin{eqnarray}
	\frac{\mathrm{d}W}{\mathrm{d}E_c} &=& \mathcal{C} (Q_{\mathrm{EC}}-E_c)^2\sqrt{1-\frac{m_\nu^2}{(Q_{\mathrm{EC}}-E_c)^2}} \nonumber \\
	&& \times \sum_\mathrm{H}\frac{B_{\mathrm{H}}\varphi_{\mathrm{H}}^2(0)\Gamma_{\mathrm{H}}}{2\pi\left[(E_c-E_{\mathrm{H}})^2+\Gamma_{\mathrm{H}}^2/4\right]}.\label{eq_Ho_Spectrum}
	\end{eqnarray}

\noindent Here, $m_{\nu}$ is the average electron neutrino mass and \QEC\ the end-point energy of the electron capture decay. The natural Lorentzian line shape of the different spectral lines H is described by their individual electron wave function at the nucleus $\varphi_{\mathrm H}^2(0)$, the exchange and overlap corrections $B_{\mathrm H}$, the line width $\Gamma_{\mathrm H}$ and the line's central energy $E_{\mathrm H}$.

This natural line shape is then convolved with the Gaussian detector response, that in turn is modified with an exponential function to account for energy losses at the low energetic side of the lines. This energy losses are caused by athermal phonons created in the absorption process, that can travel to the solid substrate without depositing their energy in the detector. This effect can be reduced by the reducing the contact area between absorber and temperature sensor, as has for example been shown in~\cite{Fleischmann2009}. For this detector only around 10\% of the events are affected by this energy loss. The energy resolution of the detector, given by the half-width of the Gaussian, was found to be 

\begin{equation}
	\Delta E_{\mathrm{FWHM}}=8.3\ \mathrm{eV}\quad .
	\end{equation}

\noindent With the energy resolution known the spectral parameters, namely the line energies $E_{\mathrm{H}}$ as well as the line widths $\Gamma_{\mathrm{H}}$ can be extracted from the measured spectrum with the fit. The resulting values are given in table~\ref{tab_energies}.

The theoretical description of the spectrum with these parameters works very well to represent the measured spectrum shown in figure~\ref{fig_Spec_full}. It only fails to describe the events around $E=450\ \mathrm{eV}$. The origin of these events is not known and different theoretical explanation are being explored. 

The difference between the experimental line energies and the electron binding energies of the daughter atom is caused by the presence of the additional $4f$-electron after the electron capture, as well as possible solid state effects, which need to be investigated.

The line widths $\Gamma_{\mathrm{H}}$ derived form the fit of the spectrum agree well with the expected values. Only the width of the NII-line seems to be much larger than expected. On the other hand does the NII line show the smallest amount of events and is heavily overlain by the low energetic tail of the NI-line. A new experiment based on optimized detectors with reduced energy loss and better energy resolution, which are developed for the next  measurement campaigns, will be able to refine the measurement of $\Gamma_{\mathrm{NII}}$.

A preliminary analysis of the end-point energy \QEC\ from the spectral fit yields:

\begin{equation}
	Q_{\mathrm{EC}} = (2.849\pm 0.005)\ \mathrm{keV}\quad ,
	\end{equation}

\noindent which is consistent with our previously reported value of $Q_{\mathrm{EC}} = (2.80\pm 0.08)\ \mathrm{keV}$~\cite{Ranitzsch2012}.

\section{Conclusion}

In this paper we have presented the first calorimetric measurement of the OI-line at 48 eV from the capture of a $5s$ electron in \Ho. This result was possible due to new data acquisition and data analysis methods based on event anti-coincidence.

The analysis of the \Ho\ spectrum obtained using the discussed data reduction allows to define the position of the lines and the intrinsic line-width with high accuracy.

The next step will be to perform a new  measurement of the \Ho\ EC spectrum with MMCs having an energy resolution $\Delta E_{\mathrm{FWHM}}$ smaller then $3\ \mathrm{eV}$ and designed to reduce the loss of high energy phonons to a negligible level. These two improvements should allow to push the energy threshold of the detector down to at least $20\ \mathrm{eV}$, which will allow an observation of the OII-line, expected around $26\ \mathrm{eV}$. The goal is also to better describe the spectrum in the region of the NI and NII line and provide more precise input to theorists to understand the origin of the shoulder above the NI-line.

The understanding of the structure present in the spectrum and therefore the physics of the EC processes is of fundamental importance for an experiment as ECHo, which aims to reach sub-eV sensitivity on the electron neutrino mass.

\bibliography{main.bbl}

\end{document}